\documentclass[12pt]{article}
\usepackage{amssymb,amsmath,amsthm,amscd,latexsym}
\usepackage{mathrsfs}
\usepackage{mathrsfs}
\usepackage{amsfonts}
\usepackage{amsmath}
\usepackage{amssymb}
\usepackage{amscd}
\usepackage{mathrsfs}
\usepackage{amssymb}
\usepackage{amsmath}
\usepackage{amsthm}
\usepackage{latexsym}
\usepackage{indentfirst}
\usepackage{enumitem}
\usepackage{anysize}
\usepackage{bbm}

\renewcommand{\paragraph}{\roman{paragraph}}
\input cyracc.def

\newcommand{\Z}{\mathbb{Z}}
\newcommand{\F}{\mathbb{F}}

\newtheorem{theorem}{Theorem}

\newtheorem{lemma}{Lemma}
\newtheorem{proposition}{Proposition}

\theoremstyle{definition}

\begin{document}
\title{\bf Double circulant self-dual and LCD codes over Galois rings
\thanks{This research is supported by National Natural Science Foundation of China (61672036),
Technology Foundation for Selected Overseas Chinese Scholar, Ministry of Personnel of China (05015133) and
the Open Research Fund of National Mobile Communications Research Laboratory, Southeast University (2015D11) and
Key projects of support program for outstanding young talents in Colleges and Universities (gxyqZD2016008).}
}
\author{
\small{Minjia Shi$^{1,2}$, Daitao Huang$^3$, Lin Sok$^3$, and Patrick Sol\'e$^4$}\\ %Patrick Sol\'e$^4$
\and \small{${}^1$School of Mathematical Sciences, Anhui University, Hefei, 230601, China}\\
\small{${}^2$Key Laboratory of Intelligent Computing \& Signal Processing,}\\
 \small{Ministry of Education, Anhui University No. 3 Feixi Road,}\\
  \small{Hefei Anhui Province 230039, P. R. China;}\\
  \small {National Mobile Communications Research Laboratory}\\
\small {Southeast University, 210096, Nanjing,  P. R. China;}\\
[-0.8ex]
\small{${}^3$ Anhui University, Hefei, 230601, P. R. China}\\
\small{${}^4$CNRS/LAGA, University of Paris 8, 2 rue de la Libert\'e, 93 526 Saint-Denis, France}
}
\date{}
\maketitle
{\bf Abstract:} { This paper investigates the existence, enumeration and asymptotic performance of self-dual and LCD double circulant codes over Galois rings
of characteristic $p^2$ and order $p^4$ with $p$ and odd prime. When $p \equiv 3 \pmod{4},$
we give an algorithm to construct a duality preserving bijective Gray map from such a Galois ring to $\Z_{p^2}^2.$
Using random coding, we obtain families of asymptotically good self-dual and LCD codes over $\Z_{p^2},$ for the metric induced by the standard $\F_p$-valued Gray maps.
}



%\begin{CJK*}{GBK}{song}\CJKtilde
\section{Introduction}
Double circulant self-dual codes over finite fields have been studied recently in \cite{AOS}.
Double circulant self-dual codes over a commutative ring can only exist if there is a square root of $-1$ over that ring \cite{LS2}. Such a root does not exist over Galois rings of even characteristic which are not fields, but does exist over Galois rings of odd characteristic and even extension degree \cite[Lemma 3.1, Lemma 3.2]{KL}. We study these codes in the case of  Galois rings of characteristic $p^2$ and size $p^4,$ for $p$ an odd prime. A recent topic related to self-dual codes is LCD codes. They are popular because of their connections with cybersecurity \cite{CG}. We also study LCD double circulant codes over the same Galois rings. For every such Galois ring we construct a duality preserving Gray map with image $\Z_{p^2}^2,$
which maps self-dual (resp. LCD) codes to self-dual (resp. LCD) codes. Note that self-dual codes over $\Z_{p^2},$ for $p$ an odd prime, have been studied by many authors
\cite{DGW,HM}. This alphabet $\Z_{p^2},$ in turn can be mapped into $\F_p^p$ by the Gray map
studied in \cite{LB}. For the two families of codes under scrutiny we give a complete enumeration formula in length $2n$ when $n$ is coprime with $p.$ This formula relies on the CRT approach to quasi-cyclic codes over rings \cite{ASS,LS2}, and
requires to count the number of solutions of certain algebraic equations over Galois rings and over their residue fields. From there, building on Artin conjecture in arithmetic progressions \cite{M}, we construct infinitely many odd primes $n$ such that $x^n-1$ has only three factors over $\F_{p^2}$ the residue class ring of the Galois ring alphabet. Depending on the congruence class of $n$ modulo $4,$ these irreducible factors
are all three self-reciprocal, or consist of $x-1,$ and one reciprocal pair. When $n$ varies in one of these two families of primes, we obtain two infinite families of double circulant codes, one self-dual, one LCD, of length $2n.$ By expurgated random coding, we derive a lower bound on the relative Hamming distance of the $\F_p$ image of both families. This shows, in particular, that both families are good.

The material is organized as follows. The next section collects the notions and notations needed in the forthcoming sections. Section 3 contains the main results on enumeration and Section 4 the main results on asymptotics. Section 5 displays some numerical examples. Section 6 concludes the article and points out some open problems.
\section{Definitions and notation}
\subsection{Some rings}
Throughout the paper, let $p$ be an odd prime.
The ring $\Z_{p^s}$ is the ring of integers modulo $p^s.$
A linear {\it code} of length $N$ over $\Z_{p^2}$ is a submodule of $\Z_{p^2}^N$. The {\it dual} $C^\bot$ and $C^{\bot_H}$
 are understood with respect to the standard inner product and Hermitian inner product, respectively. A code is {\it self-dual} if it is equal to its dual. It is {\it LCD} (linear complementary dual) if it intersects its dual trivially.
The Galois ring $GR(p^s,p^{ms})$ of order $p^{ms}$ and characteristic $p^s$ is the Galois extension of $\mathbb{Z}_{p^s}$ with degree $m$. It is a local ring, with maximal ideal $(p).$
%A {\it Galois ring $GR(4,d)$} is a quotient ring of the form $\Z_4[x]/(h)$ where $h$ is a monic polynomial which reduces to an irreducible polynomial of the same degree over $\F_2.$
The {\it Teichmuller} set $ \mathcal{T}=\{x \in GR(p^s,p^{ms})|x^{p^{m}}=x\}$ is a set of representatives of the {\it residue field $\F_{p^m}=GR(p^s,p^{ms})/(p).$} If $r\in GR(p^s,p^{ms})$, let $\widehat{r}$ denote its image in
$\F_{p^m}$ by reduction modulo $(p).$ It is known that $GR(p^s,p^{ms})=\mathcal{T}\oplus p\mathcal{T}\oplus \cdots\oplus p^{s-1}\mathcal{T} $
(base $p$ decomposition of $GR(p^s,p^{ms})$). See \cite{WZX} for background and details.
\subsection{Double circulant codes}
Denote by $R$ the ring $\frac{\mathbb{Z}_{p^2}[y]}{(h(y))}$, where $h(y)$ is a basic irreducible polynomial over $\mathbb{Z}_{p^2}$ with $\deg(h(x))=2$.
For any ring $M$, we denote by $M^*$ the set of units in $M$.
%For any $r\in R$ let $\widehat{r}= r\pmod{p}.$
Assume that $n$ is an integer with $\gcd(n,p)=1$, and
consider the code $C$ of length $2n$ over $R$ whose generator matrix has the form of
$(I_n,A),$ where $I_n$ is the identity matrix of order $n$ and $A$ is a circulant matrix over $R$.
 Note that $C$ can be viewed as a submodule of $(\frac{R[x]}{(x^n-1)})^2$, with generator $(1,a(x))$ where the $x$-expansion of $a(x)$ is the first row of $A.$

%The Galois ring $R$ of order $p^4$ and characteristic $p^2$ is the Galois extension of $\mathbb{Z}_{p^2}$ degree $2$.
For all $p$'s, we know there exists $w \in R$ such that $w^2=-1,$ by \cite{KL}. When $p \equiv 3 \pmod 4$, the polynomial $y^2+1$ is irreducible over $\F_p,$ hence over $\Z_{p^2}.$ We may write $R=\Z_{p^2}[y]/(y^2+1),$
and take $w=y.$

\subsection{Gray map}
Recall Lagrange's four-square theorem, also known as Bachet's conjecture \cite{IR}.

{\lemma \label{L} Every natural number can be represented as the sum of four integer squares.}

We will also need the sum of two squares theorem, and the Diophantus identity \cite{IR}.
{\lemma \label{STS} An integer greater than one can be written as a sum of two squares if and only if its prime decomposition contains no prime congruent to $3 \pmod{4}$ raised to an odd power.}
{\lemma \label{D} The product of two sums of two squares is a sum of two squares in following two different ways.
\begin{eqnarray}
  \left(a^{2}+b^{2}\right)\left(c^{2}+d^{2}\right) &=& \left(ac-bd\right)^{2}+\left(ad+bc\right)^{2} \\
  &=& \left(ac+bd\right)^{2}+\left(ad-bc\right)^{2}.
\end{eqnarray}
}
 Assume $p \equiv 3 \pmod{4}.$ By Lemma \ref{L}, we define a Gray map from $R$ to $\mathbb{Z}_{p^2}^2$ as follows:
$$\phi:R\rightarrow \mathbb{Z}_{p^2}^2$$
$$a+by\mapsto (ka+sb,ta+rb),$$ where $3p^2=k^2+s^2+t^2+r^2.$
Then extend it in the obvious way to $R^N.$ This Gray map is bijective, as the next result shows.
{\theorem The matrix $\begin{pmatrix} k &t\\ s & r \end{pmatrix}$ is nonsingular.  }
\begin{proof}
Suppose that, looking for a contradiction, a linear dependence between the two rows of the matrix. Let $\mu, \nu$ be two integers satisfying $\nu (k,t)=\mu (s,r).$ Substituting $k$ and $s$ by their values in $3\nu^2 p^2$ we get
$$3\nu^2 p^2=(\mu^2+\nu^2)(r^2+s^2). $$ By Diophantus identity (Lemma \ref{D})  the RHS is a sum of two squares. This contradicts Lemma \ref{STS}, because the integer $3$ times a square will always contain $3$ to an odd power in its primary factors decomposition.
\end{proof}

{\theorem Assume $p \equiv 3 \pmod{4}.$ For all codes $C$ over $R,$ we have $\phi(C)^\bot=\phi(C^\bot).$ If $C$ is a self-dual (resp. LCD) code of length $N$ over $R$, then $\phi(C)$ is self-dual (resp. LCD) of length $2N$ over $\Z_{p^2}.$}
\begin{proof}
For any vector $\mathbf{u}=(u_1,u_2,\cdots,u_N), \mathbf{v}=(v_1,v_2,\cdots,v_N) \in R^N$, where $u_i=a_{i_1}+b_{i_1}y, v_i=a_{i_2}+b_{i_2}y\in R~(i=1,2,\cdots,N).$ Suppose $\mathbf{u}\mathbf{v}=0 \pmod {p^2}$,  and noting that $p \equiv 3 \pmod{4}$, we have $y^2+1=0$ over $R$. Considering the standard inner product, we then obtain
$$\sum_{i=1}^{N}(a_{i_1}+b_{i_1}y)(a_{i_2}+b_{i_2}y)=\sum_{i=1}^{N}[(a_{i_1}a_{i_2}-b_{i_1}b_{i_2})
+(a_{i_1}b_{i_2}+a_{i_2}b_{i_1})y]=0,$$ which is equivalent to
$$\begin{cases}
\sum_{i=1}^{N}(a_{i_1}a_{i_2}-b_{i_1}b_{i_2})\equiv 0 \pmod {p^2},\\
\sum_{i=1}^{N}(a_{i_1}b_{i_2}+a_{i_2}b_{i_1})\equiv 0 \pmod {p^2}.
\end{cases}$$
We then naturally obtain
\begin{eqnarray*}
  \phi(\mathbf{u})\phi(\mathbf{v}) &=&\sum_{i=1}^{N}\phi(u_i)\phi(v_i)\\
  &=& \sum_{i=1}^{N}(ka_{i_1}+sb_{i_1},ta_{i_1}+wb_{i_1})(ka_{i_2}+sb_{i_2},ta_{i_2}+wb_{i_2}) \\
   &\equiv & \sum_{i=1}^{N}[(k^2+t^2)(a_{i_1}a_{i_2}-b_{i_1}b_{i_2})+(ks+tw)(a_{i_1}b_{i_2}+a_{i_2}b_{i_1})y] \\
   &\equiv & 0 \pmod {p^{2}}.
\end{eqnarray*}
This implies $$ \phi(C^\perp)\subseteq \phi(C)^\perp ,$$ and,
since $\phi$ is a bijection the first statement follows by showing that both sides have the same size.

 The second statement for LCD codes is as in \cite[Th. 5.2]{LCD}. The second statement for self-dual codes follows by plugging $C^\perp=C$ in the first statement.
\end{proof}

{\bf Example:} If $p=3,$ then $3p^2=27=16+9+1+1.$ The Gray map can be taken to be $a+by \mapsto (4a+3b,a+b).$
\subsection{Finite fields}
If $L,\,K$ are two finite fields of respective orders $p^{rs}$ and $p^r$ satisfying $K\subseteq L,$ we write the {\it trace} from $L$ down to $K$ as
$$Tr_{p^r}^{p^{rs}}(z)=z+z^{p^r}+\dots+z^{p^{r(s-1)}}, $$
where $r,s$ are positive integers.

\subsection{Codes over fields and asymptotics}
Let $p$ be an odd prime, and denote by $\F_p$ the finite field of order $p.$
By a {\bf code} of length $N$ over $\F_p,$ we shall mean a proper subset of $\F_p^N.$ This code is {\bf linear} if it is a  $\F_p$-vector subspace of $\F_p^N.$
 The {\bf dimension} of a code $C$, denoted by $k$, is equal to its dimension as a vector space.
Its (minimum) {\bf distance}, denoted by $d$ or $d(C),$ is defined as the minimum Hamming weight of its nonzero elements. The {\bf Hamming weight}  of $x=(x_1,x_2,\cdots,x_n)\in \F_p^n,$ denoted by $w(x),$ is the number of indices $i$ where $x_i \neq 0.$
The three parameters of a code are written compactly as $[n,k,d].$
We extend this notation to a possibly nonlinear code $C \subseteq \F_p^n,$ by letting then $k =\log_p(\vert C\vert),$ and letting $d$ be the minimum pairwise distance between two nonzero codewords.
If $C(n)$ is a family of codes of parameters $[n, k_n, d_n]$, the {\bf rate} $r$ and {\bf relative distance} $\delta$ are defined as $$r=\limsup\limits_{n \rightarrow \infty}\frac{k_n}{n},$$ and
\begin{equation*}\delta=\liminf\limits_{n \rightarrow \infty}\frac{d_n}{n}.\end{equation*}
A family of codes is said to be {\bf good} iff $r\delta >0.$

Recall that the $q$-ary {\bf entropy function} $H_q(\cdot)$ is defined for $0<y< \frac{q-1}{q}$ by $$ H_q(y)=y\log_q(q-1)-y\log_q(y)-(1-y)\log_q(1-y).$$

\section{Main results}

%There is a natural notion of conjugation on $R$ induced by the complex conjugation. Let $z=t+\alpha t'$ be a generic $z\in R$ with $t,t' \in \mathbb{Z}_{p^2}$. We shall define by $\overline{z}$ the conjugate of $z$ and define it as $\overline{z}=t+\overline{\alpha}t'=t-\alpha t'$. We then define the trace of $z\in R$ down to $\mathbb{Z}_{p^2}$ by $T(z)=z+\overline{z'}$.

\subsection{Enumeration in a  special case}
Assume that $p$ is a primitive root modulo $n$ with $n$ an odd prime. Then, ${(p^2)}^{\frac{n-1}{2}}\equiv 1 \pmod{n},$ and we have $x^n-1=(x-1)\tilde{g_2}(x)\tilde{g}_3(x)$
with $\tilde{g_2}(x),\tilde{g_3}(x)$ monic irreducible polynomials over $\mathbb{F}_{p^2}$.
By Hensel lifting, since $(n,p)=1,$ we have
$x^n-1=\vartheta(x-1){g_2}(x){g}_3(x)$ over $R,$ with $g_2(x) \equiv \tilde{g_2}(x) \pmod{p}$, $g_3(x) \equiv \tilde{g_3}(x) \pmod{p}$ and
$\vartheta \in R^*$. Both $g_2(x)$ and $g_3(x)$ are monic basic irreducible polynomials over $R.$
The following lemma is taken from \cite{WZX}.
{\lemma Let $R_1=GR(p^s,p^{sm})$ and $h(x)$ be a monic basic irreducible polynomial of degree $l$ over $R_1$. Then the residue class ring $\frac{R_1[x]}{(h(x))}$ is a Galois ring of characteristic $p^s$ and cardinality $p^{sml}$ and contains $R_1$ as a subring. Thus $$\frac{R_1[x]}{(h(x))}=GR(p^s,p^{sml}).$$}\vspace{-0.3cm}
%Write $\xi=x+(h(x))$, then $h(\xi)=0$, all elements of $\frac{R[x]}{(h(x))}$ can be expressed uniquely in the form $$a_0+a_1\xi+\cdots+a_{l-1}\xi^{l-1},$$ where $a_0,a_1,\cdots,a_{l-1}\in R$, and $\frac{R[x]}{(h(x))}=R[\xi].$}

Thus this lemma shows that the alphabet rings ${\mathcal R}_i$ for $i=2,3$ of the constitutent codes defined below are also Galois rings.

\subsubsection{$n \equiv 1 \pmod 4$}
Firstly, if $n \equiv 1 \pmod 4$, since $p$ is primitive modulo $n,$ we can deduce $({p^2})^{\frac{n-1}{4}} \equiv -1 \pmod n.$ This implies that $-1$ is in the $p^2$-cyclotomic coset modulo $n.$ Consequently, we obtain $g_i^*(x)=g_i(x)$ for $i=2,3.$
 Note that $\deg(g_i(x))=\frac{n-1}{2}$. By the CRT, we have $C=C_1\bigoplus C_2\bigoplus C_3$, where $C_1$ is a code of length $2$ over $R$ and
 $C_i$ is a code of length $2$ over the ring $\frac{R[x]}{(g_i(x))}$ for $i=2,3$.  Let $\mathcal{R}_i=\frac{R[x]}{(g_i(x))}$ for $i=2,3.$
 The properties of $C_3$ being similar to that of $C_2$, we only investigate $C_2$ for simplicity's sake.

Write $C_2=\langle[1,b]\rangle$, where $b\in \mathcal{R}_2.$ Note that $b$ can be uniquely decomposed in base $p$ as $b=\alpha + p\beta$,
where $\alpha,\beta \in \mathcal{T}=\{x \in \mathcal{R}_2~|~x^{p^{n-1}}=x\}$.
Define a generalized {\bf Frobenius map} $F$ as $F(b)=\alpha^{p^2}+p\beta^{p^{2}}$, then
$F^{\frac{n-1}{4}}(b)=\alpha^{p^{\frac{n-1}{2}}}+p\beta^{p^{\frac{n-1}{2}}}$.
The {\bf conjugate} $\overline{b}$ of $b$ is $F^{\frac{n-1}{4}}(b)$. Throughout this paper, we let $u=p^{\frac{n-1}{2}}$. With this notation $\overline{b}=\alpha^u+p \beta^u.$
We can then define a {\bf Hermitian scalar product} on $\mathcal{R}_2^2$, by the formula $x\cdot \overline{y}=x_1\overline{y_1}+x_2\overline{y_2}$
for $x=(x_1,x_2),y=(y_1,y_2)\in \mathcal{R}_2^2.$

From the introduction we know there is $w\in R$ such that $w^2=-1.$ Write $w=w_1+pw_2$, $w_1,w_2 \in \mathcal{T}_1=\{x\in R \mid x^{p^2}=x\}.$
Then $w^2=w_1^2+2pw_1w_2=-1,$ which is equivalent to $w_1^2=-1, w_2=0$. That is to say, only has two choices for $w$ in $R$ such that $w^2=-1$.
Let us denote one of this two $w'$s by $\sqrt{-1}.$
 {\theorem \label{c1sd} Writing $C_1=\langle[1,a]\rangle$, where $a \in R$. Then we have $C_1$ is self-dual over $R$ iff $a= \pm \sqrt{-1}$.}
  \begin{proof} $C_1$ is self-dual iff $1+aa=0$, which implies $a= \pm \sqrt{-1}$.
\end{proof}

The following lemma generalizes the {\it Yamada normal form}  of \cite{WZX}.
{\lemma \label{111} Let $A,B\in \mathcal{T}$, then we have $A+B=T_1+pT_2$ with $T_1, T_2 \in \mathcal{T}$ is uniquely given by
$$\begin{cases}
T_1=(A^{\frac{1}{p}}+B^{\frac{1}{p}})^p,\\
T_2 \equiv-\mathcal{P}_{p}(A^{\frac{1}{p}},B^{\frac{1}{p}})\pmod p,
\end{cases}$$ where $p\mathcal{P}_p(A,B)=\sum_{i=1}^{p-1}(_i^p)A^iB^{p-i}, $ and $\mathcal{P}_p(A,B)$ is a polynomial in $A,B$ with integral coefficients.}
\begin{proof} Note that $$(A+B)^p=A^p+B^p+p\mathcal{P}_p(A,B).$$
We then claim that $(A+B)^{p^i}=A^{p^{i}}+B^{p^{i}}+p\mathcal{P}_p(A^{p^{i-1}},B^{p^{i-1}})$ for any integer $i.$
This can be proved by induction on $n$ as follows.
\begin{eqnarray*}
  (A+B)^{p^n} &=& (A^{p^{n-1}}+B^{p^{n-1}}+p\mathcal{P}_p(A^{p^{n-1}},B^{p^{n-1}}))^p \\
   &=&(A^{p^{n-1}}+B^{p^{n-1}})^p\\
   &=& A^{p^{n}}+B^{p^{n}}+p\mathcal{P}_p(A^{p^{n-1}},B^{p^{n-1}}),
    \end{eqnarray*}
where the first and last equality hold by induction hypothesis.
We then obtain $$A+B=(A^{\frac{1}{p}}+B^{\frac{1}{p}})^p-p\mathcal{P}_p(A^{\frac{1}{p}},B^{\frac{1}{p}}),$$ which implies
$$\begin{cases}
T_1=(A^{\frac{1}{p}}+B^{\frac{1}{p}})^p,\\
T_2 \equiv-\mathcal{P}_{p}(A^{\frac{1}{p}},B^{\frac{1}{p}})\pmod p,
\end{cases}$$
since it can be checked by the formulas above that $(A^{\frac{1}{p}}+B^{\frac{1}{p}})^{p^n}=(A^{\frac{1}{p}}+B^{\frac{1}{p}})^p,$ showing that $(A^{\frac{1}{p}}+B^{\frac{1}{p}})^p \in {\mathcal T}.$ This determines $T_1$ uniquely.
We can only determine $T_2 \pmod p,$ which is enough for our purpose.
Then the result follows.
\end{proof}

With Lemma 5, we then obtain the following important theorem, which gives a necessary and sufficient condition for $C_2$ to be a self-dual code
over $\mathcal{R}_2$.
{\theorem $C_2$ is self-dual over $\mathcal{R}_2$ with respect to Hermitian inner product iff $$\begin{cases}
 1+\alpha^{\frac{1+u}{p}}\equiv 0 \pmod p,\\
\beta\alpha^u+\beta^u\alpha-\mathcal{P}_{p}(1,\alpha^{\frac{1+u}{p}})\equiv 0 \pmod p.
\end{cases}$$}
\begin{proof}
$C_2$ is self-dual iff $1+b\overline{b}=0$, i.e.
$1+bF^{\frac{n-1}{4}}(b)=0,$ which is equivalent to
\begin{eqnarray}
    & 1+\alpha^{1+u}+p\beta\alpha^u+p\beta^u\alpha=0=T_1+pT_2,&
      \end{eqnarray}
      where $T_1\in \mathcal{T}, T_2\in \mathcal{T}$ and $T_1=T_2=0.$

 By Equation (3) and Lemma 5, we can then have $$1+\alpha^{1+u}=(1+\alpha^{\frac{1+u}{p}})^p-p\mathcal{P}_p(1,\alpha^{\frac{1+u}{p}}).$$ Then Equation $(3)$ is equivalent to $$(1+\alpha^{\frac{1+u}{p}})^p+p(\beta\alpha^u+\beta^u\alpha-\mathcal{P}_{p}(1,\alpha^{\frac{1+u}{p}}))=0.$$
 That implies $$\begin{cases}
 1+\alpha^{\frac{1+u}{p}}\equiv 0 \pmod p\\
\beta\alpha^u+\beta^u\alpha-\mathcal{P}_{p}(1,\alpha^{\frac{1+u}{p}})\equiv 0 \pmod p
\end{cases}$$Then the result follows.
\end{proof}

Next, we will enumerate the number of possible choices for $C_2.$

{\theorem \label{sdc2} The number of self-dual codes $C_2$ over $\mathcal{R}_2$ is equal to $u(1+u).$}
\begin{proof}
Let $x=\alpha^{\frac{1}{p}}$ and we then consider the equation $$x^{1+u}\equiv -1 \pmod p.$$ It can be obtained that, noting that $2(1+u)$ is a divisor of $u^2-1$,
the number of choices for $x,$ hence for $\alpha,$ is $1+u.$
Then by the equation $\beta\alpha^u+\beta^u\alpha-\mathcal{P}_p(1,\alpha^{\frac{1+u}{p}})\equiv 0 \pmod p$, we get
$$Tr_u^{u^2}(\widehat{\beta}\widehat{\alpha}^u)=f(\widehat{\alpha}),$$ where $f(\widehat{\alpha})$ denotes $\mathcal{P}_p(1,\alpha^{\frac{1+u}{p}})\equiv 0 \pmod p.$
That implies that there are $u$ choices for $\beta$ when fixed $\alpha$.
Thus, the number of $C_2$ is equal to $u(1+u).$
\end{proof}

By the CRT, the following theorem can be derived.

{\theorem The number of self-dual codes $C$ over $R$ is equal to $2u^2(u+1)^2.$ }
\begin{proof}
By Theorems \ref{c1sd} and \ref{sdc2}, we obtain the number of $C$'s as $2u^2(u+1)^2.$
\end{proof}

We are now ready to investigate the number of LCD codes. We firstly introduce the following proposition.

{\proposition $C_2$ is LCD over $\mathcal{R}_2$ iff $1+b \cdot \overline{b} \in \mathcal{R}_2^*$.}
\begin{proof}
$``\Leftarrow"$  \begin{enumerate}
\item If $b \in \mathcal{R}_2^*$, we then obtain $C_2^{\bot_H}=\langle[1,-\frac{1}{\overline{b}}]\rangle$. Suppose $C_2$ is not LCD, then there exist nonzero $t,k \in \mathcal{R}_2$ such that $t(1,b)=k(1,-\frac{1}{\overline{b}})$. We then obtain $t(1+b\overline{b})=0$, which implies $t=0$ since $1+b \cdot \overline{b} \in \mathcal{R}_2^*$. Contradiction !
\item If $b \in \mathcal{R}_2\backslash \mathcal{R}_2^*$, write $b=pb'$ with $b' \in \mathcal{T}$. Then we have \begin{eqnarray*}
                                                C_2^{\bot_H}&=& \left(
      \begin{array}{cc}
        2\overline{b'} & 1 \\
        0 & 2b'' \\
      \end{array}
    \right)
                                              \end{eqnarray*}

    with $b''\in \mathcal{T}$. Suppose $C_2$ is not LCD, then there exist nonzero $m,n,l\in \mathcal{R}_2$ such that $m(1,2b')=n(2\overline{b'},1)+l(0,2b'')$. That implies
    $$\begin{cases}
    m=2n\overline{b'},\\
    2mb'=n+2lb'',
    \end{cases}$$
    then we obtain $m=0$, a contradiction !
\end{enumerate}
$``\Rightarrow"$ Suppose $1+b\cdot \overline{b}\in \mathcal{R}_2\backslash \mathcal{R}_2^*$, then we have the following two cases.
\begin{enumerate}
  \item If $b \in \mathcal{R}_2^*$, we have $p(1+b\cdot \overline{b})=0=p(1,b)\cdot\overline{(1,b)}$. Note that  $C_2^{\bot_H}=\langle[1,-\frac{1}{\overline{b}}]\rangle$, we then obtain $p(1,b)\in C_2^{\bot_H}$, which implies $p(1,b)\in C_2^{\bot_H}\cap C_2.$ Contradiction!
  \item If $b \in \mathcal{R}_2\backslash \mathcal{R}_2^*$, we then get $1\in \mathcal{R}_2\backslash \mathcal{R}_2^*$, a contradiction !
\end{enumerate}

This completes the proof. \end{proof}

Similarly, we have Proposition 2 about $C_1$, which is a constituent code over $R$.

{\proposition The code $C_1$ is LCD over $R$ iff $1+a \cdot a \in R^*$. In particular, the number of possible choices for $a$ is $p^4-2p^2$.}
\begin{proof}
The proof of the first statement is similar to that of Proposition $1$, with the Euclidean inner product replacing the Hermitian inner product. We omit it here. Write $a=a_1+pa_2$ with $a_1,a_2 \in \mathcal{T}_1$.
By this criterion the number of possible choices for $a$ is equal to $$p^4-|\{a \mid a^2+1=pa', a'\in \mathcal{T}_1\}|=p^4-|\{(a_1,a_2)\mid a_1^2+1=0\}|,$$ where $\mathcal{T}_1=\{x\in R\mid x^{p^2}=x\}$. Then the result follows.
\end{proof}

To determine the number of LCD codes $C_2$, we will reason by complementation.

{\theorem $C_2$ is not LCD code over $\mathcal{R}_2$ with respect to Hermitian inner product iff $$
 1+\alpha^{\frac{1+u}{p}}\equiv 0 \pmod p.
$$}\vspace{-0.8cm}
\begin{proof}
$C_2$ is  not LCD iff $1+b\overline{b}\in p\mathcal{R}_2$, i.e.,
\begin{eqnarray}
    & 1+\alpha^{1+u}+p\beta\alpha^u+p\beta^u\alpha=T_1+pT_2,&
      \end{eqnarray}
      where $T_1=0, T_2\in \mathcal{T}$.
 By Equation (4) and Lemma 5, we can then have $$1+\alpha^{1+u}=(1+\alpha^{\frac{1+u}{p}})^p-p\mathcal{P}_p(1,\alpha^{\frac{1+u}{p}}).$$ Then Equation $(4)$ is equivalent to $$(1+\alpha^{\frac{1+u}{p}})^p+p(\beta\alpha^u+\beta^u\alpha-\mathcal{P}_{p}(1,\alpha^{\frac{1+u}{p}}))=pT_2.$$
 That implies $$
 1+\alpha^{\frac{1+u}{p}}\equiv 0 \pmod p,$$
with $\beta$ is arbitrary.
\end{proof}

Combining Proposition 2 and Theorem 7, we have the following important theorem.

{\theorem \label{8}The number of LCD codes $C$ over $R$ is equal to $(p^4-2p^2)(p^{2(n-1)}-u^3-u^2)^2$.}
\begin{proof}
We know that the total number of codes over $R$ and $\mathcal{R}_2$ are $p^4$ and $p^{2(n-1)}$, respectively. Based on Theorems \ref{c1sd} and \ref{sdc2}, the choice for non LCD codes is $(u+1)u^2$, then the result follows by subtraction.
\end{proof}

\subsubsection{$n \equiv 3 \pmod 4$}

If $n \equiv 3 \pmod 4$, we have $g_3^*(x)=g_2(x),g_2^*(x)=g_3(x).$ Then by the CRT, we obtain $C=C_1\bigoplus C_2'\bigoplus C_3'$, where $C_2'$ is a code of length $2$ over the ring $\frac{R[x]}{(g_2(x))}$ and $C_3'$ is a code of length $2$ over the ring $\frac{R[x]}{(g_2^*(x))}$. Writing $C_2'=\langle[1,b']\rangle$. Let $\mathcal{R}_i'=\frac{R[x]}{(g_i(x))}$ for $i=2,3.$ Similarly, $b'$ can be uniquely decomposed in base $p$ as $b'=\alpha' + p\beta'$, where $\alpha',\beta' \in \mathcal{T}'=\{x \in \mathcal{R}_2'~|~x^{p^{n-1}}=x\}$.

Note that, if $C$ is self-dual we then have $C_3'=C_2'^\bot$. The following result is needed to count the number of self-dual $C$'s.

{\theorem \label{c2sd} The number of dual pairs $(C_2',C_3'=C_2'^\bot)$ over $\mathcal{R}_2'$ is equal to $p^{2(n-1)}-p^{n-1}$.}
\begin{proof}
Assume that $C_2'=\langle[1,b']\rangle$, let $C_3'=C_2'^{\bot}$, then we have the following discussion.
\begin{enumerate}
  \item [(i)]If $b'$ is a unit, we then obtain $C_3'=\langle[1,-\frac{1}{b'}]\rangle$.

  \item [(ii)]If $b'$ is not a unit, let $b'=p\beta'$, then we get $C_3'=\langle[-p\beta',1]\rangle$.

\end{enumerate}
Thus, from the form of the generator matrix of $C_3'$, it is clear that the number of dual pairs $(C_2',C_3'=C_2'^\bot)$ is exactly the size of $\mathcal{R}_2'^{*}$, i.e., is equal to $p^{2(n-1)}-p^{n-1}$.
\end{proof}
{\theorem The number of self-dual codes $C$ over $R$ is equal to $2\cdot (p^{2(n-1)}-p^{n-1}).$ }
\begin{proof}
Based on Theorems \ref{c1sd} and \ref{c2sd}, we can obtain the desired results.
\end{proof}
{\lemma \label{3} Writing $C_2'=\langle[1,b']\rangle$ and $C_3'=\langle[1,c']\rangle$. Then $$\begin{cases}
C_2'\cap C_3'^\bot=\{0\},\\
C_2'^\bot\cap C_3'=\{0\}
\end{cases}$$ iff $1+b'c'\notin p\mathcal{R}_2^{'}.$}
\begin{proof}
$``\Rightarrow"$ If $1+b'c' \in p\mathcal{R}_2^{'}$, we then obtain $p(1+b'c')=0$, which implies $p(1,b')\cdot(1,c')=0$. That is equivalent to $p(1,b') \in C_3'^\bot$, noting that $p(1,b') \in C_2'$, which is a contradiction with $C_2'\cap C_3'^\perp=\{0\}$.

$``\Leftarrow"$ If $C_2'\cap C_3'^\perp \neq\{0\}$, then we must have $\lambda \in GR(p^2,p^{2(n-1)})$ such that $\lambda (1,b')(1,c')=\lambda(1+b'c')=0$.
\begin{enumerate}
  \item [(1)] If $\lambda$ is a unit, we get $1+b'c'=0$, which is a contradiction with $1+b'c' \notin p\mathcal{R}_2'.$
  \item [(2)] If $\lambda$ is not a unit, writing $\lambda=p\lambda_1$ with $\lambda_1 \in \mathcal{T}_{e_j}$, we can obtain $1+b'c' \in p\mathcal{R}_2',$  Contradiction !
\end{enumerate}

Then the result follows.\end{proof}
{\theorem \label{10} Suppose $u'=p^{n-1}$, the number of LCD codes over $R$ is equal to $(p^4-2p^2)\cdot(u'^4-u'^2+u').$}
\begin{proof}
Just keep the same notations as in Lemma \ref{3}. In the following, we aim to count the possible choices for $C_2'$ and $C_3'$. Then we have
\begin{enumerate}
  \item [(1)] If $b'$ is a unit, then $c'$ is arbitrary except the case when $c'\in \frac{-1}{b'}+p\mathcal{R}_2'$, which implies the number of pairs $(b',c')$ is $(u'^2-u')^2$.
  \item [(2)] If $b'$ is not a unit, i.e., $b'=pb'_1$ with $b'_1\in \mathcal{T}'$, then $c'$ is arbitrary. In detail, $c'=c'_1+pc'_2$ with $c'_1,c'_2 \in \mathcal{T}'$,
  we have $1+b'c'=1+pc'_1b'_1 \notin p\mathcal{R}_{2}'.$ Thus, the number of pairs $(b',c')$ is $u'\cdot u'^2=u'^3.$
\end{enumerate}
Thus, the total number of $C_j',C_j''$ is $(u'^2-u')^2+u'^3=u'^4-u'^3+u'^2$. Therefore, the total number of LCD double circulant codes is equal to, based on Proposition 2, $$(p^4-2p^2)\cdot(u'^4-u'^3+u'^2).$$
This completes the proof.\end{proof}

\subsection{Enumeration in the general case}
The following result, while not needed for the asymptotics, is of interest in its own right.
{\theorem Let $n$ be an odd integer, assume that the factorization of $x^n-1$ into irreducible polynomials over $R$ is of the form $$x^n-1=\varsigma (x-1)\prod_{i=2}^sg_i(x)\prod_{j=1}^th_j(x)h_j^*(x),$$ with $\varsigma \in R^*$, and $g_i(x)$ is a self-reciprocal polynomial of degree $d_i$ with $d_i$ a even integer, the polynomial $h_j(x)$ is of degree $e_j$ and $*$ denotes reciprocation. The number of self-dual double circulant codes over $R$ is $$2\prod_{i=2}^{s}(u_i^2+u_i)\prod_{j=1}^{t}(u_j'^2-u_j'),$$
The number of LCD double circulant codes over $R$ is $$(p^4-2p^2)\prod_{i=2}^{s}(u_i^4-u_i^3-u_i^2)\prod_{j=1}^{t}(u_j'^4-u_j'^3+u_j'^2),$$ where $u_i=p^{d_i}, u_j'=p^{2e_j}.$  }
\begin{proof}
Let $\mathcal{R}=\frac{R[x]}{(x^n-1)}$.
We know that $$\mathcal{R}\simeq \frac{R[x]}{(x-1)}\oplus \bigg(\bigoplus_{i=2}^s\frac{R[x]}{(g_i(x))}\bigg)\oplus \bigg(\bigoplus_{j=1}^t(\frac{R[x]}{(h_j(x))}\oplus\frac{R[x]}{(h_j^*(x))})\bigg)$$ by the CRT.
Denote by $G_i$ the ring $\frac{R[x]}{(g_i(x))}$,  by $H_j'$ the ring $\frac{R[x]}{(h_j(x))}$ and by $H_j''$ the ring $\frac{R[x]}{(h_j^*(x))}$. This decomposition naturally extends to $\mathcal{R}^2$ as $$\mathcal{R}^2\simeq R^2\oplus \bigg(\bigoplus_{i=2}^s G_i^2\bigg)\oplus \bigg(\bigoplus_{j=1}^t (H_j'^2\oplus H_j''^2)\bigg).$$
In particular, each $\mathcal{R}$-linear code of length 2 can be decomposed as the ``CRT sum"
$$C\simeq C_1\oplus \bigg(\bigoplus_{i=2}^s C_i\bigg)\oplus \bigg(\bigoplus_{j=1}^t (C_j'\oplus C_j'')\bigg).$$
By Theorem \ref{c1sd}, and the analogues of Theorems \ref{sdc2} and \ref{c2sd}, the number of self-dual codes is $$2\prod_{i=2}^{s}u_i(1+u_i)\prod_{j=1}^{t}(u_j'^2-u_j').$$ Based on Theorems \ref{8} and \ref{10}, the number of LCD codes is equal to
$$(p^4-2p^2)\prod_{i=2}^{s}(p^{4d_i}-u_i^2(u_i+1))\prod_{j=1}^{t}(u_j'^4-u_j'^3+u_j'^2),$$
where $u_i=p^{d_i}, u_j'=p^{2e_j}.$ Then the result follows.
\end{proof}

\section{Relative distance bound}
This section only uses enumeration for the special case of the factorization of $x^n-1$ into two irreducibles. We assume $p\equiv 3 \pmod 4$ to use the duality-preserving Gray map given by
Section 2.3.
If $C$ is an $R$-code, we call its {\em $\F_p$ image} the image of $\phi(C)$ by the Gray map of \cite{LB} with $k=1.$
Note that if $C$ is a code of length $2n$ over $R$ then $\phi(C)$ is of length $4n$ over $\Z_{p^2}$ and the $\F_p$-image of $\phi(C)$   has length $4pn$ over $\F_p.$ We prepare for the proof of the main result by
the following result.
{\theorem \label{lambda}  If $e,f\in R^n,$ and $(0,0) \neq (e,f)$ has Hamming weight $<n$, then the vector $(e,f)$ is in at most $\lambda$ double circulant codes of length $2n$ with $\lambda=p^{3n+1}.$ }
 \begin{proof}
  Write $(e,f)=(e_1,f_1)\oplus (e_2,f_2)\oplus (e_3,f_3)$ for the CRT decomposition of $(e,f).$
  Consider $C_1=\langle[1,a]\rangle, a=\alpha_1+p\beta_1, \alpha_1,\beta_1\in \mathcal{T}_1=\{x\in R~|~x^{p^2}=x\}.$ Let $(e_1,f_1)\in C_1$, we have $f_1=e_1a$.
  \begin{enumerate}
    \item [(i)]If $e_1$ is a unit, $a=\frac{f_1}{e_1}$.
    \item [(ii)]If $e_1(\neq 0)$ is not a unit, write $e_1=pe_1'$ with $e_1'\in \mathcal{T}_1^*$. We then obtain $f_1=pae_1'$.

  \begin{enumerate}
  \item [(ii-1)]$\alpha_{1}\neq0,$ we have $f_1=pf_1'=p\alpha_{1} e_1'$, which implies $\alpha_{1}=\frac{f_1'}{e_1'}$, $\beta_{1}$ is arbitrary.
  \item [(ii-2)]$\alpha_{1}=0$, we have $a=p\beta_{1}$, which implies $f_1=0$, $\beta_{1}$ is arbitrary.
  \end{enumerate}
   \item [(iii)]If $e_1=0$, we then get $f_1=0$. That implies $a$ is arbitrary. We can easily get there are at most $p^4$ double circulant codes that containing $(e_1,f_1)$.
  \end{enumerate}
   Consider $C_2=\langle[1,b]\rangle, b=\alpha_2+p\beta_2, \alpha_2,\beta_2\in \mathcal{T}_2$, where the case is similar to that of $C_1$ with  $\mathcal{T}_2$ playing the role of $\mathcal{T}_1$. Note that $\mathcal{T}_2=\mathcal{T}$ or $\mathcal{T}'$.
  Thus, the number of constituent codes $C_2$ is at most $|\mathcal{T}|^2=|\mathcal{T}'|^2$.

The case is the same as that of $C_3.$ However, these two analyses cannot be run independently. The detail is as follows.
\begin{enumerate}
  \item If both $e_2$ and $e_3$ are $0$, then we obtain $e\in ({g_2(x)g_3(x)}),$ the repetition code of length $n.$ So either $e=0$ yielding $f=0,$ or
 $w_H(e)=n,$ contradicting the hypothesis. This argument shows that the case (iii) cannot happen simultaneously for $C_2$ and $C_3.$
  \item If both $e_2$ and $e_3$ are not units, and $e_2,e_3 \neq 0$, we can easily get $\widehat{e} \equiv \widehat{e_2e_3}\equiv 0 \pmod {g_2(x)g_3(x)}$.
       \begin{enumerate}
         \item  If $\widehat{e} \neq 0$, we then obtain $w_H(e)\geq w_H(\widehat{e})=n$, a contradiction !
         \item If $\widehat{e} = 0 $, assume that $0 \neq e=pe'$, then we have $f=pf'$. Write $C=\langle[1,d]\rangle$, where $d=d_1+pd_2$ with $d \in \frac{R[x]}{(x^n-1)}, d_1,d_2\in \{x\in \frac{R[x]}{(x^n-1)} \mid x^{p^{2n}}=x\}$. We can get $d_1=\frac{f'}{e'}$ and $d_2$ is arbitrary. In this case, there are at most $p^{2n}$ double circulant codes containing $(e,f)$.
       \end{enumerate}

\end{enumerate}

  Thus, we obtain $\lambda=p^4|\mathcal{T}|^{2}|\mathcal{T}|=p^{3n+1}$ when $e_1=e_2=0$ and $e_3$ is not unit.
\end{proof}
{\theorem \label{lambda1}  If $e,f\in R^n,$ and $(0,0) \neq (e,f)$ has Hamming weight $<n$, then the vector $(e,f)$ is in at most $\lambda$ self-dual double circulant codes of length $2n$ with $\lambda=2u^2(u+1)$. }
 \begin{proof}
  Write $(e,f)=(e_1,f_1)\oplus (e_2,f_2)\oplus (e_3,f_3)$ for the CRT decomposition of $(e,f).$
  Let $(e_1,f_1)\in C_1=\langle[1,a]\rangle, a=\alpha_1+p\beta_1, \alpha_1,\beta_1\in \mathcal{T}_1=\{x\in R~|~x^{p^2}=x\}.$ We can easily get there at most exist $2$ self-dual codes which contain $(e_1,f_1)$. Consider $(e_2,f_2)\in C_2=\langle[1,b]\rangle,~b=\alpha_2+p\beta_2$, with $\alpha_2,\beta_2\in \mathcal{T}$ or $\mathcal{T}'$, we then obtain $f_2=be_2$.
  \begin{enumerate}
    \item [(i)]If $e_2$ is a unit, then $b=\frac{f_2}{e_2}$.
    \item [(ii)]If $e_2(\neq 0)$ is not a unit, write $e_2=pe_2'$ with $e_2'\in \mathcal{T}_2^{*}=\{x\in R~|~x^{p^2}=x\}$. We then obtain $f_2=pae_2'$.
  \begin{enumerate}
  \item [(ii-1)]$\alpha_{2}\neq0,$ we have $f_2=pf_2'=p\alpha_{2} e_2'$, which implies $\alpha_{2}=\frac{f_2'}{e_2'}$, $\beta_{2}$ is arbitrary.
  \item [(ii-2)]$\alpha_{2}=0$, we have $b=p\beta_{2}$, which implies $f_2=0$, $\beta_{2}$ is arbitrary.
  \end{enumerate}
  \item [(iii)]If $e_2=0$, we then get $f_2=0$. That implies $b$ is arbitrary. We can easily get there are at most $u(1+u)$ self-dual codes that containing $(e_2,f_2)$.
  \end{enumerate}

By Theorem \ref{lambda}, then $e_2,e_3\neq 0$.
If $e_2,e_3$ are not units with $e_2,e_3\neq 0$, based on $2$ in Theorem \ref{lambda}, for the subcase $2(b)$, we obtain $1+d^2=0$.
We then get $1+d_1^2+2pd_1d_2=0$, where $d=d_1+pd_2$. Then we have $1+d_1^2\equiv 0 \pmod p$, which implies two coices for $d_1.$
By Lemma \ref{111}, we can get $2\widehat{d_1}\widehat{d_2}-\mathcal{P}_p(1,\widehat{d_1}^{\frac{2}{p}})=0,$ which implies
$\widehat{d_2}=\frac{\mathcal{P}_p(1,\widehat{d_1}^{\frac{2}{p}})}{2\widehat{d_1}}$. Then $d_2$ is uniquely determined by $d_1$.
That is to say, there are at most $2$ self-dual codes containing $(e,f)$ when $e_2,e_3$ are not units.

 Thus, by Theorem $5$, we obtain $\lambda=2u^2(u+1)$  with at most $u$ self-dual codes containing $(e_3,f_3)$ if $n\equiv 1 \pmod 4$. When $n\equiv 3 \pmod 4$, $C_3$ is determined by $C_2$.

Then, the result follows.\end{proof}

We can now state and prove the main result of this paper.
{\theorem Assume the Artin conjecture for primes in arithmetic progression \cite{M} holds. There is an infinite family of double circulant self-dual (resp. LCD) $R$-codes with rate $\frac{1}{2}$ and relative Hamming distance of the $\F_p$ image $\delta \ge H_p^{-1}(1/8p)$ (resp. $\delta \ge H_p^{-1}(1/4p)$).   }
\begin{proof} Artin conjecture for primes in arithmetic progression shows in particular that $p$ and $\epsilon \in \{\pm 1\}$ being given, there are infinitely many primes $n\equiv \epsilon \pmod{4}$ such that $p$ is primitive modulo $n.$
 It should be noted that the size $\Omega_n$ of the family of codes we consider is asymptotically equivalent to $ 2u^4$ for self-dual codes and to $p^{4n}$ for LCD codes. This holds for the case $\epsilon=1$ by Subsection 3.1.1
 like for the case $\epsilon=-1$ by Subsection 3.1.2.
Assume we can prove that for $n$ large enough
$\Omega_n>\lambda B(d_n)$, where $\lambda=p^{3n+1}$ for LCD codes and $\lambda=2u^2(u+1)$ for self-dual codes and $B(r)$ denotes the number of vectors in $R^{2n}$ with Hamming weight of their $\F_p$ image $<r.$
This would imply by Theorem \ref{lambda} that there are codes of length $2n$ in the family with $\F_p$ image distance $\ge d_n.$ Denote by $\delta$ the relative distance of this family
of $p$-ary codes. If we take $d_n$ the largest number satisfying the said inequality, and assume a growth of the form
$d_n\sim 4p \delta_0 n,$ then, using an entropic estimate for $B(d_n)\sim p^{4pnH_p(\delta_0)}$ (cf. \cite[Lemma 2.10.3]{W}) yields, with the said values of $\Omega_n$ and $\lambda$ the estimate
  $H_p(\delta_0)=\frac{1}{8p}$ for self-dual codes and $H_p(\delta_0)=\frac{1}{4p}$ for LCD codes. The result follows by observing that, by definition of the family of codes so constructed, $\delta\ge \delta_0.$
\end{proof}

\section{Numerical examples}
Assume $p=3$ and $C=\langle [I,A_0+yA_1]\rangle,$ where $A_0,\,A_1$ are circulant matrices of size $n$ and $I$ is the identity matrix of the same size. The generator matrix of $\phi(C)$ can be computed as

$$\begin{pmatrix} 4I&I&4A_0+3A_1&A_0+A_1\\3I&I&3A_0-4A_1&A_0-A_1 \end{pmatrix}.$$
Define the base $3$ decomposition of $x\in \mathbb{Z}_9$ as $$x=r_0(x)+3r_1(x),$$ where $r_i(x)\in \{0,1,2\}$. Then we can define the Gray map of \cite{LB} as
$$\Phi: \mathbb{Z}_9 \rightarrow \mathbb{Z}_3^3$$
$$a \mapsto (a_0,a_1,a_2),$$ where $a_i=r_1(a)+ir_0(a)$.
The explicit map is tabulated below.\\
\begin{center}
\begin{tabular}{|c|c|c|}
\hline
$x$ & $\Phi(x)$ & $w_H(\Phi(x))$\\
  \hline
  0 & (0,0,0) & 0 \\
  1 & (0,1,2)& 2\\
  2 & (0,2,1)& 2 \\
  3 & (1,1,1)& 3\\
  4 & (1,2,0) &2\\
  5 & (1,0,2)&2 \\
  6 & (2,2,2) &3 \\
  7 & (2,0,1)& 2\\
  8 & (2,1,0)&2 \\
  \hline
\end{tabular}
\end{center}

In Table \ref{table:LCD} and Table \ref{table:SD} we have collected some examples of double circulant LCD and self-dual codes obtained by random search in Magma. The coefficients of degree $n$ polynomial $a(x)$ are written in decreasing powers of $x,$ for example for
$n=3,$ the entry $811$ means $8x^2+x+1.$
The parameters over $\Z_9$ and $\Z_3$ are given in the form $(4n,9^{2n},d_{\phi(C)})$ and $[12n,4n,d_{\Phi(C)}]$ respectively, where $d_{\phi(C)}$ and $d_{\Phi(C)}$ are the Hamming minimum distancess of their Gray images $\phi$ and $\Phi$ respectively.
The entry in the rightmost column is the best known distance of an $[12n,4n]$ tenary linear code, obtained by looking up at the tables in {\tt www.codetables.de}.
\begin{table}\caption{Double circulant LCD codes over $\Z_9+u\Z_9$ and their Gray images $\phi, \Phi$}
$$
\begin{array}{|c|l|l|c|c|c|}
\hline
n&\textbf{$a_1(x)$}&\textbf{$a_0(x)$}&(4n,9^{2n},d_{\phi(C)})_{ \Z_9}&[12n,4n,d_{\Phi(C)}]_{ \Z_3}&\textbf{Distance of BKLC over } \Z_3\\
\hline
 2& 41 &51&(8 ,9^4 ,4 )&[24,8,6]&11\\
 3& 8 1 1 &0 8 1&(12, 9^6, 6)&[36, 12, 12]&15\\
4& 3 6 5 1 &6 5 0 5&(16, 9^8, 5)&[48, 16, 14]&18\\
 5& 1 0 8 5 6  &5 7 6 6 4&(20, 9^{10}, 6)&[60, 20, 16]&21\\
\hline
\end{array}
$$\label{table:LCD}
\end{table}

\begin{table}\caption{Double circulant self-dual codes over $\Z_9+u\Z_9$ and their Gray images $\phi, \Phi$}
$$
\begin{array}{|c|l|l|c|c|c|}
\hline
n&\textbf{$a_1(x)$}&\textbf{$a_0(x)$}&(4n,9^{2n},d_{\phi(C)})_{ \Z_9}&[12n,4n,d_{\Phi(C)}]_{ \Z_3}&\textbf{Distance of BKLC over } \Z_3\\
\hline
 2& 10 &00&(8 ,9^4 ,3 )&[24,8,10]&11\\
 3& 8 1 1 &0 8 1&(12, 9^6, 6)&[36, 12, 12]&15\\
% 4& 4 0 4 0&8 8 1 1&(16, 9^8, 6)&[48, 16, 12]&18\\
4& 6731&4752&(16, 9^8, 6)&[48, 16, 15]&18\\
 5&26758  &62532&(20, 9^{10}, 6)&[60, 20, 18]&21\\
\hline
\end{array}
$$\label{table:SD}
\end{table}
\section{Conclusion}
In this article we have studied double circulant codes either self-dual or LCD over Galois rings of characteristic $p^2$ and size $p^4.$ Extending the study to $GR(p^s,p^{ms}),$ with $s>2$ would result in more terms in the base $p$ expansion of a ring element and would make the computations of Section 3 more difficult. A similar remark can be made about using $m>2.$ More tractable could be to study quasi-cyclic codes of higher index, like four-circulant codes, for instance.

We have used the composition of two Gray maps to derive codes over $\F_p.$ While the choice of the Hamming metric over $\F_p$ is the most natural one, the study of the Lee minimum distance of the $\F_p$-image could also be worthwhile.


\begin{thebibliography}{99}
\bibitem{AOS}A. Alahmadi, F. \"Ozdemir, P. Sol\'e, On self-dual double circulant code, {\it Designs, Codes Cryptogr.}, online July 20, 2017.
\bibitem{CG} C. Carlet, S. Guilley, Complementary Dual Codes for Counter-Measures to Side-Channel Attacks,
 Adv. in Math. of Comm. , {\bf 10}(1), (2016), 131-150 .
\bibitem{DGW} S.T. Dougherty, T.A. Gulliver, and J. Wong, Self-Dual Codes over $\Z_8$ and $\Z_9,$ Designs, Codes, and Cryptography,(2006), 235--249.
\bibitem{LCD} S. T. Dougherty, J.L. Kim, B. \"Ozkaya, L. Sok, P. Sol\'e,
The combinatorics of LCD codes: linear programming bound and orthogonal matrices. Int. J. of Information and Coding Theory, {\bf 4}(2/3), (2017), 116--128.
\bibitem{HM} M. Harada, A. Munemasa, On the classification of self-dual $\Z_k$-codes,Lecture Notes in Comput. Sci.,5921, Springer, (2009), 78--90.
\bibitem{W} W. C. Huffman, V. Pless, {\it Fundamentals of error correcting codes}, Cambridge University Press, 2003.
\bibitem{IR}K. Ireland, M. Rosen, {\it A Classical Introduction to Modern Number Theory }(2nd ed.) (1990), Springer
\bibitem{KL}J-L. Kim, Y. Lee,
Construction of MDS self-dual codes over Galois rings. Des. Codes Cryptography, {\bf 45}(2), (2007), 247--258.
\bibitem{LB} S. Ling, T. Blackford, $\Z_{p^{k+1}}$-linear codes, IEEE trans. on Information Theory {\bf 48}(9), (2002), 2592--2605.
\bibitem{LS2}S. Ling, P. Sol\'e, On the Algebraic Structure of Quasi-cyclic Codes II: Chain Rings, Des. Codes Cryptography, {\bf 30}(1), (2003), 113--130.
\bibitem{M}P. Moree, On primes in arithmetic progression having a prescribed primitive root, Journal of Number Theory, {\bf 78}(1), (1999), 85-98.
\bibitem{ASS}M. Shi, A. Alahmadi, P. Sol\'e,{\it  Codes and Rings: Theory and Practice}, Academic Press (2017).

%\bibitem{OYD} Z. Odemis-Ozger, B. Yildiz, S. T. Dougherty, On Codes over $\Z_{p^s}$ with the Extended Lee Weight,Filomat,{\bf 30:2} (2016), 255�C268
\bibitem{WZX} Z. X. Wan, {\it Finite Fields and Galois Rings,} World Scientific (2003).

\end{thebibliography}
\end{document}